\documentclass[12pt]{article}

\begin{document}
\title{Ellipsoidal shapes in general relativity: \\general definitions and
an application}

\author{%
J\'ozsef Zsigrai\thanks{E-mail: zsigrai@sunserv.kfki.hu}\\
  {\small \it Department of Physics, Hiroshima University,}\\
  {\small \it Higashi-Hiroshima 739-8526, Japan} \\
  {\small \it and}\\
  {\small \it Institute for Isotope- and Surface Chemistry of the Hungarian
Academy of Science}\\
{\small \it H-1525 Budapest, P.O. Box 77, Hungary} }

\maketitle

\renewcommand{\theequation}{\thesection.\arabic{equation}}

\newcommand{\be}{\begin{equation}}
\newcommand{\ee}{\end{equation}}

\begin{abstract}
A generalization of the notion of ellipsoids to curved Riemannian
spaces is given and the possibility to use it in  describing the shapes of rotating bodies
in general relativity is examined. As an illustrative example, stationary,
axisymmetric perfect-fluid spacetimes with a so-called confocal inside
ellipsoidal symmetry are investigated in detail under the assumption that the
4-velocity of the fluid is parallel to a time-like Killing vector field. A
class of perfect-fluid metrics representing interior NUT-spacetimes is obtained
along with a vacuum solution with a non-zero cosmological constant.


\end{abstract}

\section{Introduction}

In the past much effort has been made to find  an exact solution of Einstein's field equations representing the gravitational field of a compact rapidly rotating object, but without much success. In particular, a great deal of work has been done concerning stationary, axially symmetric perfect-fluid configurations as possible candidates for modelling the interior of rapidly rotating bodies. In addition, in order to give a global space-time model completely describing the gravitational field of the source, the possibilities for matching an interior perfect-fluid solution to an exterior vacuum metric have been extensively studied. So far, however, the only known physically acceptable global exact solution in the examined category is the so-called Neugebauer-Meinel disk of dust \cite{NM}, which may be considered as the limiting case of uniformly rotating relativistic objects. 

Apart from the above mentioned disk, in many cases encountered in the search for physically acceptable solutions it is not possible to give an obvious geometrical interpretation for the discovered configurations. On the other hand, it is well known that in Newtonian gravity numerous results have been obtained for the case in which a rotating fluid in equilibrium has an ellipsoidal shape \cite{Chandra}.  In the present work we will constrain our attention to those rapidly rotating configurations in general relativity, for which there is a natural way for imposing the restriction that the surface of the considered object is ellipsoidal. Using the approach presented in this paper the matching surface between the interior and exterior region may be conveniently chosen to be the ellipsoid representing the boundary of the rotating fluid body. It should be emphasized, however, that other shapes for a rotating fluid might also be possible in both general relativity and Newtonian theory as well. 

In the Euclidean geometry of flat space the notion of an ellipsoid is clear and easily understood. In curved spaces, however, it does not have such an obvious intuitive meaning. Therefore, in the general-relativistic analysis of perfect-fluid bodies having an ellipsoidal shape, it is necessary to determine first what an ellipsoid in curved space is. Originally, it has been proposed by Krasi\'nski \cite{Krasinski} to consider a 3-dimensional, axially symmetric Riemannian space in which the line element $^{(3)}ds^2$ of the 3-dimensional space can be written as  
\be
^{(3)}ds^2=z(\rho,\theta)d\rho^2+(\rho^2+a^2\cos^2\theta)d\theta^2+(\rho^2+a^2)
\sin^2\theta d \phi^2 , \label{dsconfocalrotinv}
\ee
where $a$ is a constant and $z$ is a positive function. In such spaces the
coordinate surfaces $\rho=const$ represent the analogue of a one-parameter
congruence of confocal Euclidean ellipsoids of revolution. Later a more general,
coordinate-free definition of (not necessarily axially symmetric) Riemannian
spaces that can be filled in with a one-parameter family of confocal ellipsoids
was given by R\'acz \cite{RaczCQG}, recovering the metric (\ref{dsconfocalrotinv}) as a
special case. In \cite{Krasinski} it was assumed that the union of local rest spaces of the observers co-moving with matter has the metric (\ref{dsconfocalrotinv}), whereas in \cite{RaczCQG} it was the 3-space of timelike Killing orbits  which had this ellipsoidal symmetry.  By using the approach of \cite{RaczCQG}, a uniqueness theorem for stationary axisymmetric vacuum spacetimes with a particular type of ellipsoidal symmetry was proven in \cite{RaczSuveges}. We remark here that ellipsoidal shapes in curved spaces, but in a different context, have been considered in \cite{Vacaru} as well (see also references therein).

The purpose of the present paper is to construct a framework for considering ellipsoidal shapes in general relativity, which covers the situations studied in \cite{Krasinski} and \cite{RaczCQG} as special cases. In this way a firm mathematical basis for the ideas pioneered by Krasi\'nski \cite{Krasinski} is given and the constructions introduced  by R\'acz \cite{RaczCQG} are extended in order to include a wider class of spaces in the analysis. This work also aims to examine the applicability of the constructed framework on the particular example of stationary, axially symmetric rigidly rotating perfect fluid spacetimes with certain ellipsoidal symmetries as test models for the interior solution. It has been shown by Geroch and Lindblom \cite{GerochLindblom} that rigidly rotating perfect fluids can be considered to be the equilibrium configurations of relativistic dissipative fluids so they are of obvious astrophysical interest. Although the test models considered in this paper did not provide perfect-fluid solutions with physically suitable properties, the existence of such solutions within the very wide class of spacetimes with some kind of ellipsoidal symmetry cannot be excluded.

This paper is organized as follows. First the definition of a Riemannian 3-space
that can be filled in with a one-parameter congruence of concentric ellipsoids
is given. In such curved spaces one can introduce a notion
analogous to the notion of ellipsoids in Euclidean
space. Next, the possibilities for considering this type of 3-spaces
within the framework of general relativity are discussed. The definition of a
``conformally ellipsoidal spacetime'' as well as the more restrictive
definitions of a ``spacetime with co-moving ellipsoidal symmetry'' and of a
``spacetime with inside ellipsoidal symmetry'' are presented. As an illustration
of these notions, stationary, axially symmetric, rigidly-rotating perfect-fluid
configurations with ``confocal'' inside ellipsoidal symmetry are considered.
After reviewing some basic notions pertaining to rigidly rotating fluids, our
considerations are first specialized to the subcase in which the gradient of the
norm of the time-like Killing vector field and its twist are linearly dependent.
Making use of the integrability conditions of the field equations, the general
form of the metric for the selected perfect-fluid configurations is determined and it is
demonstrated that the resulting metric can be joined smoothly to the Taub-NUT
solution. Some previously known exact solutions are shown to belong to the class
of  spacetimes studied here. In the complementary case when the norm of the
gradient and the twist of the Killing field are linearly independent it is shown
that a particular (possibly new) perfect-fluid solution satisfies the
integrability conditions. The resulting particular metric, however, can be
re-interpreted as a vacuum solution with a non-zero cosmological constant.
Finally, the Appendix contains a review of some facts concerning ellipsoids in
flat Euclidean space.

\section{``Ellipsoids'' in curved space}\label{ECS}
\setcounter{equation}{0}

In the following a possible approach for considering ellipsoidal shapes in curved Riemannian spaces is presented. In particular, the definition given in \cite{RaczCQG} is generalized and it is shown how a convenient coordinate system can be chosen in the Riemannian spaces under consideration.

\smallskip
\noindent {\bf Definition 1}: A Riemannian 3-space (${\cal S}, h_{ab}$) {\it can
be filled in with a one-parameter congruence of concentric
ellipsoids} if it satisfies the following properties: (a) there exists in it a
congruence of surfaces, $\cal E_\rho$, with induced metric $\chi_{ij}[\rho]$,
(b) there exists a diffeomorphism, $\Psi$, mapping $\cal S$ into a subset,
$\Psi [{\cal S}]$, of $\bf R^3$ so that the tensor field $\Psi^*\chi_{ij}[\rho]$
is just the induced metric field for some one-parameter congruence of concentric
ellipsoids in $\Psi[{\cal S}]\subset {\bf R^3}$ and (c)   for any value of
$\rho$, the extrinsic curvature tensor field, $\kappa_{ij}[\rho]$, of the
surfaces $\cal E_\rho$, embedded in $\cal S$, is conformal to the metric tensor
of the Euclidean unit sphere, $\gamma_{ij}$. \smallskip

We will refer to the surfaces $\cal E_\rho$ as ``ellipsoids in curved space''.
Note that this definition has been obtained by generalizing  the properties of
Euclidean ellipsoids (given in the Appendix) to curved spaces, just as it has
been done in \cite{RaczCQG}. However, an essential difference between the above
definition and the one given in \cite{RaczCQG} is that the term ``confocal''
used in \cite{RaczCQG} is replaced here by the term ``concentric''. This allows
one to consider Riemannian spaces of a much more general character than it was
the case in \cite{RaczCQG}.

By analogy with the 3+1 decomposition of a spacetime (${\cal M},
g_{ab}$)  (see, e.g., \S 21.4 of \cite{MTW}), a 3-dimensional Riemannian space
(${\cal S}, h_{ab}$) may be decomposed in a similar way in 2+1 dimensions. Let
$\cal E_\rho$ denote a surface labelled by the parameter value ``$\rho$'' in a
one-parameter congruence of 2-dimensional surfaces embedded in a 3-dimensional
Riemannian space. Since there is always only one surface $\cal E_\rho$ of the
congruence passing through any given point of ${\cal S}$, the parameter $\rho$
used for labelling the surfaces may be chosen as one of the coordinates of $\cal
S$. In such a coordinate system the distance between two nearby points
belonging  to two different surfaces $\cal E_\rho$ and ${\cal E}_{\rho+d\rho}$
can be given as \be
ds^2=\alpha^2d\rho^2+\chi_{ij}[\rho](dx^i+\beta^i d\rho)(dx^j+\beta^j d\rho) ,
\label{2+1} \ee
where $\alpha$ and $\beta^i$ are the 2+1 analogues of the
 ``lapse function'' and ``shift vector'', $\chi_{ij}[\rho]$ is the
induced metric on $\cal E_\rho$ , while
$x^i$  ($i,j=2,3$) is a coordinate system on this surface.  Expression
(\ref{2+1}) gives, in fact,  the line element of the 3-dimensional Riemannian
space (${\cal S}, h_{ab}$). The components of the metric $h_{ab}$ and its
inverse, $h^{ab}$, are given by

\begin{eqnarray}
\left(\begin{array}{cc}
     h_{11} &  h_{1j} \\
     h_{i1} & h_{ij} \\
     \end{array}\right)
     &=&
  \left( \begin{array}{cc}
     \alpha^2+\beta_k\beta^k &  \beta_j \\
     \beta_i & \chi_{ij} \\
     \end{array} \right), \label{habandinverseA}\\
\left(\begin{array}{cc}
     h^{11} &  h^{1j} \\
     h^{i1} & h^{ij} \\
     \end{array}\right)
     &=&
  \left( \begin{array}{cc}
     1/\alpha^2 &  -\beta^j/ \alpha^2 \\
     -\beta^i/\alpha^2 & \chi^{ij}+\beta^i\beta^j/\alpha^2 \\
     \end{array} \right) .   \label{habandinverseB}
\end{eqnarray}

The requirement that $\Psi^*\chi_{ij}[\rho]$ is the induced metric field for
some one-para\-me\-ter congruence of concentric ellipsoids in $\Psi[{\cal
S}]\subset {\bf R^3}$ implies that there exists a coordinate system
$(x^2=\theta, x^3=\phi)$ in which the line element of the induced metric on the
surface $\cal E_\rho$ can be written as (compare with equation
(\ref{dsEuclideanEllipsoidal}) of the Appendix)
\begin{eqnarray}
ds^2(\chi_{ij}[\rho])&=&
\biggl[(x_0^2\cos^2\phi+y_0^2\sin^2\phi)\cos^2\theta+
\rho^2\sin^2\theta\biggr]\cdot d\theta^2 \nonumber \\
&+&(x_0^2\sin^2\phi+y_0^2\cos^2\phi)\sin^2\theta \cdot d\phi^2 \! \ , \label{chi}\\
&+&2(y_0^2-x_0^2)\sin\theta\cos\theta\sin\phi\cos\phi \cdot d\theta d\phi \nonumber
\end{eqnarray}
where $x_0=x_0(\rho)$ and $y_0=y_0(\rho)$.
On the other hand, the requirement that the extrinsic curvature tensor field
$\kappa_{ij}[\rho]$ is conformal to the metric of the Euclidean unit sphere,
$\gamma_{ij}$, means that there exists a smooth (at least $C^2$)
everywhere positive function $k(\rho,\theta,\phi)$ so that
\footnote{From now on we omit the labels
``$[\rho]$'' in the notation and denote any tensor field $\tau_{ij}[\rho]$
simply as $\tau_{ij}$. Thus, the dependence on $\rho$, as well as on the
coordinates $\theta$ and $\phi$, is understood.}  \be
\kappa_{ij}=k(\rho,\theta,\phi)\gamma_{ij} \label{extrinsiccurv} \ . \ee

In general, the extrinsic curvature, or ``second fundamental form'', of a
surface is given by the expression  \cite{DyerOliwa, MusgraveLake}
\be
\kappa_{ij}={{\partial x^a}\over{\partial\xi^i}}
 {{\partial x^b}\over{\partial\xi^j}}D_\alpha n_b  \ , \label{ExtrinsicCurvDef}
 \ee
 where $n_a$ is the unit normal to the surface, $D_a$ is the
covariant derivative associated with the metric of the Riemannian 3-space,
$x^a$ are coordinates on the 3-space, while  $\xi^i$ are coordinates
on the 2-dimensional surface ($a,b=1,2,3$ and $i,j=2,3$). If the surface is
given by an equation of the form
\be f(x^a(\xi^i))=0
\ee
then the unit normal can be given as
\be
n_a=\pm \biggl| h^{bc}
{{\partial f}\over{\partial x^b}}
{{\partial f}\over{\partial x^c}}\biggr|^{-1/2}
{{\partial f}\over{\partial x^a}}
\ee
with the ambiguity in sign arising from the ambiguity in the orientation of the
normal. In the present case
$x^1=:\rho$, $x^2=\xi^2=:\theta$ $x^3=\xi^3=:\phi$, the surface equation is
given by $f=\rho-const=0$, and, thus, the components of the unit normal are
$(n_1, n_2, n_3)=(\pm \alpha, 0, 0)$.

In this setup, with the aid of
(\ref{habandinverseA}) and  (\ref{habandinverseB}) one obtains from
(\ref{ExtrinsicCurvDef})  \begin{equation}
\kappa_{ij}=\pm {1\over {2\alpha}}\biggl(
{{\partial\beta_j}\over{\partial x^i}}+{{\partial\beta_i}\over{\partial x^j}}-
{{\partial\chi_{ij}}\over{\partial \rho}}-2\beta_k \tilde\Gamma_{ij}^k
\biggr)
\label{kappacomponents}
\end{equation}
for the extrinsic curvature (see also \cite{MTW}), where $\tilde \Gamma_{ij}^k$
are the Christoffel symbols for the metric $\chi_{ij}$, which is given by
(\ref{chi}). Furthermore, in the coordinate system of (\ref{chi}), the
line element of the unit sphere may be written as
\be
ds^2(\gamma_{ij})=d\theta^2+\sin^2\theta d\phi^2 ,
\label{gammaij}
\ee
so it
follows from equation (\ref{extrinsiccurv}) that
\begin{eqnarray}
{{\kappa_{(\phi\phi)}}\over{\kappa_{(\theta\theta)}}}=\sin^2\theta, \
\kappa_{(\theta\phi)}=0, \label{kappasystem}
\end{eqnarray}
where $\kappa_{(\theta\theta)}$, $\kappa_{(\theta\phi)}$ and
$\kappa_{(\phi\phi)}$ each denote the corresponding component of the extrinsic
curvature. Since for the spaces under consideration the metric $\chi_{ij}$ has
been already specified by equation (\ref{chi}), it follows from
(\ref{kappacomponents}) that the system (\ref{kappasystem}) contains as unknown
functions only the two components of $\beta_i$.

Now it is straightforward to show that  (\ref{kappasystem}) is identically
satisfied if one choses the components $\beta_{(\theta)}$ and $\beta_{(\phi)}$
of $\beta_i$ as
 \begin{eqnarray}
\beta_{(\theta)}&=&\bigl[(x_0x_0'\cos^2\phi+y_0y_0'\sin^2\phi-\rho)\cos
\theta+F(\rho)\bigr]\sin\theta , \\
\beta_{(\phi)}&=&\bigl[(-x_0x_0'+y_0y_0')\sin\phi\cos\phi
+G(\rho)\bigl]\sin^2\theta ,
\end{eqnarray}
where $F(\rho)$ and $G(\rho)$ are arbitrary functions
and the prime denotes derivation with respect to $\rho$. Note that
$\beta_{(\theta)}$ and $\beta_{(\phi)}$ satisfy the system (\ref{kappasystem})
for {\it arbitrary} choice of the functions $F$ and $G$, including the particular
case $F=G=0$.  In fact, fixing the functions $F$ and $G$ merely corresponds to
a definite choice of coordinate system. Therefore, without loss of
generality, hereafter these two functions will be set to be identically zero.
Correspondingly, there exists a coordinate system in which the line element of a curved Riemannian 3-space that can be
filled in with a one-parameter congruence of concentric ellipsoids takes the
form
\begin{eqnarray}
 ds^2&=&z(\rho,\theta,\phi)\cdot d\rho^2
+\bigl[(x_0^2\cos^2\phi+y_0^2\sin^2\phi)\cos^2\theta+
\rho^2\sin^2\theta\bigr]\cdot d\theta^2 \nonumber\\
 &+&(x_0^2\sin^2\phi+y_0^2\cos^2\phi)\sin^2\theta \cdot d\phi^2 \nonumber\\
 &+&2(x_0x_0'\cos^2\phi+y_0y_0'\sin^2\phi-\rho)\cos\theta
\sin\theta \cdot d\rho d\theta \  , \label{dsEllipsoidal}\\
 &+&2(-x_0x_0'+y_0y_0')\cos\phi\sin\phi
\sin^2\theta \cdot d\rho d\phi \nonumber\\
 &+&2(y_0^2-x_0^2)\cos\phi\sin\phi\cos\theta\sin\theta \cdot d\theta
d\phi \nonumber
\end{eqnarray}
where $z(\rho,\theta,\phi)$ is a $C^2$ function, $x_0$ and $y_0$ are functions
depending merely on the coordinate `$\rho$', and the prime denotes derivation.
In these coordinates the $\rho=const$ surfaces are the curved-space analogues of
Euclidean ellipsoids. Note also that the particular choice
$z=((x_0')^2\cos^2\phi+(y_0')^2\sin^2\phi)\sin^2\theta + \cos^2\theta$ recovers
the metric of flat Euclidean space in ellipsoidal coordinates (compare with
equation (\ref{dsEuclideanEllipsoidal}) of the Appendix).

The freedom carried by the functions $x_0$ and $y_0$ as well as
the $\phi$-dependence of $z$ in the above form of the metric implies that
(\ref{dsEllipsoidal}) describes  a much wider class of spaces than the metric
(\ref{dsconfocalrotinv}). Putting, e.g., $x_0^2=\rho^2+a^2$,
$y_0^2=\rho^2+b^2$ into (\ref{dsEllipsoidal}) where $a$ and $b$ are
constants, recovers the line element of non-symmetric Riemannian spaces that can
be filled in with a congruence of {\it confocal} ellipsoids, originally given in
\cite{RaczCQG}\footnote{Note, however, that the non-diagonal term with ``$d\theta
d\phi$'' is missing form the line element given by equation (11) in
\cite{RaczCQG}. Nevertheless, this omission does not influence the remaining
results in \cite{RaczCQG}, because they correspond to the rotationally
invariant case, in which the mentioned non-diagonal term identically vanishes.}.
The present paper, however, will be particularly concerned with the special case
in which $z(\rho,\theta,\phi)=z(\rho,\theta)$ and $x_0^2=y_0^2=\rho^2+a^2$,
where $ a=const$. Then the metric of the considered 3-dimensional Riemannian
space reduces exactly to the metric (\ref{dsconfocalrotinv}). In this special
case the congruence of the $\rho=const$ surfaces is the analogue of a congruence
of rotationally invariant confocal oblate ellipsoids in Euclidean space.

\section{Ellipsoidal shapes in general relativity}
\setcounter{equation}{0}

In the previous section we have been concerned merely with structures in
3-dimensional Riemannian geometry, without connecting them to any structures in
general relativity. In this section Riemannian spaces within the 4-dimensional
spacetime of general relativity will be considered, assuming that these
Riemannian spaces can be filled in with a one-parameter family of concentric
ellipsoids, in the sense of Definition 1 from the previous section.

In general relativity a description of any shape depends on the motion of the
observers performing the description, as it has been already emphasized in
\cite{Krasinski}. E.g., a body that has  an ellipsoidal shape as seen by one
class of observers might appear to have a completely different shape as seen by
some other observers. Therefore when considering ellipsoidal shapes in general
relativity, one should also specify a choice of preferred observers. To start
off, consider the following construction:

\smallskip
\noindent {\bf Definition 2}: A spacetime  $(M, g_{ab})$ is {\it conformally
ellipsoidal} if there exists in it a congruence of preferred observers for which
the union of their local rest spaces is conformal to a Riemannian 3-space
$({\cal S}, h_{ab})$ that can be filled in with a one-parameter congruence of
concentric ellipsoids.
\smallskip

For the observers moving with 4-velocity $u^a$, Definition 2 implies
\be
g_{ab}+u_au_b=\Omega^2 h_{ab} ,\label{definition}
\ee
where $\Omega^2$ is a conformal factor. A spacetime is {\it ellipsoidal} if
$\Omega^2=1$. Note that the original term ``ellipsoidal
spacetime'' first introduced in \cite{Krasinski} is equivalent to the notion of
a stationary, axisymmetric spacetime with ``co-moving ellipsoidal symmetry''
(see Definition 3) in the terminology of the present work.

Notice now that the choice of preferred observers has still not been uniquely
specified. Therefore, in accordance with the remark above Definition 2, in
order to consider ellipsoidal shapes within general relativity, Definition
2 should be supplemented by a prescription for choosing the preferred
observers. There is an enormous freedom in fixing the congruence
of observers and the conformal factor referred to in Definition 2. In
particular, we will consider two possible situations, which give rise to the
definitions presented bellow, generalizing the constructions given in
\cite{Krasinski} and \cite{RaczCQG}.

Let us recall that the main motivation for considering the analogues of
ellipsoidal surfaces in curved spaces was examining rapidly rotating bodies
having an ellipsoidal surface in the framework of general relativity. As
discussed in \cite{Krasinski}, a natural 3-dimensional Riemannian space in which
one should consider the ellipsoidal shapes of rotating bodies in general
relativity is the union of local rest spaces of observers co-moving with matter.
This serves as a motivation for Definition 3.

\smallskip
\noindent {\bf Definition 3}: A non-empty spacetime possesses {\it co-moving
ellipsoidal symmetry}  if it is ellipsoidal (i.e., conformally ellipsoidal with
$\Omega^2=1$) and if the class of preferred observers is exactly the class of
observers co-moving with matter. Furthermore, any ellipsoidal vacuum spacetime
will be said to possess co-moving ellipsoidal symmetry, by definition.
\smallskip

In an empty spacetime, in general, there is no way to specify a class of
distinguished observers. However, if there is a Killing vector field
present on spacetime, one may give, for instance, the following definition,
regardless whether the spacetime is empty or non-empty:

\smallskip
\noindent {\bf Definition 4}:  A spacetime with a timelike Killing vector field
$\xi^a$ possesses {\it inside ellipsoidal symmetry} if it is conformally
ellipsoidal with $\Omega^2=(-\xi_a\xi^a)^{-1}=:(-v)^{-1}$ and if the
4-velocity of the preferred observers, $u^a$, is a unit vector field aligned
with the Killing vector field, i.e., if $u^a=(-v)^{-1/2}\xi^a$.
\smallskip

In particular, if the rest spaces of the preferred observers from Definition 4
can be filled in with a congruence of {\it confocal} ellipsoids, we will refer
to the spacetime as possessing {\it confocal inside ellipsoidal
symmetry}\footnote{We remark that the term ``inside ellipsoidal symmetry''
originally introduced in \cite{RaczCQG} is equivalent to the term ``confocal
inside ellipsoidal symmetry'' used in the present work.}.

A spacetime with co-moving ellipsoidal symmetry need not, but might possess
inside ellipsoidal symmetry and vice versa. For instance, the Kerr spacetime is
an example of a spacetime which possesses both these types of
ellipsoidal symmetry at the same time. The present paper will be mainly
concerned with the case of inside ellipsoidal symmetry, while spacetimes with
co-moving ellipsoidal symmetry are the subject of ongoing research. It turns
out, however, that a class of metrics with inside ellipsoidal symmetry discussed
in this paper possesses co-moving ellipsoidal symmetry as well (see equations
(\ref{insideNUTlike}) and (\ref{NUTlike}) later on in this paper).

In \cite{RaczSuveges} stationary, axially symmetric empty spacetimes with
confocal inside ellipsoidal symmetry have been considered\footnote{Note that
\cite{RaczSuveges} refers to ``ellipsoidal spacetimes'' which are, in fact,
spacetimes with confocal inside ellipsoidal symmetry in the terminology of
the present paper.}. By examining the integrability conditions of the field
equations it has been demonstrated that the full set of solutions to the vacuum
Einstein equations satisfying these conditions coincides with the family of the
Kerr-NUT solutions with vanishing electric and magnetic charges. In the
remaining part of the present paper a method similar to that of
\cite{RaczSuveges} will be used to study stationary, axisymmetric perfect-fluid
configurations possessing confocal inside ellipsoidal symmetry and having the
4-velocity of the fluid parallel to a timelike Killing vector field.

\section{Rigidly rotating perfect-fluid spacetimes
with confocal inside ellipsoidal symmetry}
\setcounter{equation}{0}

\subsection{The field equations for rigidly rotating perfect fluids}

In the following we will restrict our attention to stationary, axisymmetric
perfect-fluid spacetimes with the energy momentum tensor
\be
T_{ab}=(\mu+P)u_a u_b+Pg_{ab}\ , \label{Tab}\ee
where $\mu$ and $P$ are the density and pressure of the fluid and for which the
4-velocity of the fluid is a unit vector field $u^a$ aligned with a timelike
Killing field $\xi^a$, so that
\be u^a=(-v)^{-1/2}\xi^a ,\ee
where $v=\xi_a\xi^a$ is the norm of the time-like Killing vector field $\xi^a$.
Such fluids are ``rigid'' in the sense that they are expansion- and shear-free.
Moreover, for the selected configurations the twist of the Killing field,
defined by $\omega_a:=\varepsilon_{abcd}\xi^b\nabla^c\xi^d$, can be given as a
gradient of a function, that is, $\omega_a=\nabla_a \omega$, where $\omega$ is
called the twist potential. The existence of the Killing field allows one to use
the projection formalism of general relativity (for a review see, e.g., \cite{Exact, Geroch, ZsigraiPTP, Maison}) and to write the field equations for the
selected perfect-fluid configurations in the form \cite{RaczJMP,RaczERE}
\begin{equation}R_{ab}-16\pi
v^{-1}Ph_{ab}={1\over 2}v^{-2}\bigl[ (D_av)(D_bv)+(D_a\omega)(D_b\omega)\bigr]\
, \label{fe1} \end{equation}
\begin{equation}D_mD^mv=v^{-1}\bigl[(D_mv)(D^mv)-(D_m\omega)(D^m\omega)
\bigr]-8\pi (\mu+3P)\ ,\label{fe2} \end{equation}
\begin{equation}D_mD^m\omega=2v^{-1}(D_m\omega)(D^mv)\ , \label{fe3}
\end{equation} where $R_{ab}$ and $D_a$ are the Ricci-tensor and covariant
derivative associated with the 3-dimensional Riemannian metric $h_{ab}$ on the
space of time-like Killing orbits defined by  \begin{equation}
(-v)^{-1}h_{ab}=g_{ab}-{1\over v}\xi_a\xi_b. \label{hab}
\end{equation}
In addition to the above equations, the equation of motion
\begin{equation} \partial _aP+{1\over 2} (\mu+P){{\partial _av}\over
v}=0\ \label{EL},\end{equation}
which  follows from the above ones, is also satisfied.

In the spirit of Definition 4, for spacetimes with confocal inside ellipsoidal
symmetry the 4-velocity of the preferred observers is given by
$u^a=(-v)^{-1/2}\xi^a$, i.e., in the present case of rigidly rotating fluids it
coincides with the 4-velocity of the fluid elements. Therefore, the preferred
observers referred to in Definition 4 are now co-moving with the fluid. This,
however,  does not necessarily mean that the spacetime possesses co-moving
ellipsoidal symmetry in the sense of Definition 3, because that would
also require $\Omega^2=1$ for the conformal factor $\Omega^2$ introduced in
equation (\ref{definition}). In the present case $\Omega^2=(-v)^{-1}$ and
$h_{ab}$ is expected to be the metric of an axially symmetric Riemannian 3-space
that can be filled in with a one-parameter congruence of confocal ellipsoids.
Thus, equation (\ref{definition}) is equivalent now to equation (\ref{hab}),
which defines the metric on the space of timelike Killing orbits. Using
(\ref{dsconfocalrotinv}) for the line element of the metric $h_{ab}$, it follows
that in a coordinate system $(x^0=t, x^1=\rho, x^2=\theta,
x^3=\phi)$\footnote{Note that, in general, the coordinates $x^1$, $x^2$ and
$x^3$ of spacetime may depend on the coordinates $\rho$, $\theta$ and $\phi$ of
the three-dimensional Riemannian space in a more complicated way. In the present
case of rigidly rotating fluids with inside ellipsoidal symmetry, however, the
simple identifications $ x^1=\rho, x^2=\theta, x^3=\phi$ are possible.} adapted
to the Killing vector fields, the line element of a stationary, axially
symmetric spacetime possessing confocal inside ellipsoidal symmetry can be given
in the form
 \begin{eqnarray}
 ds^2&=&
v(\rho,\theta)\bigl(dt+A(\rho,\theta)d\phi\bigr)^2 \nonumber \\
 &-&{1\over {v(\rho,\theta)}} \biggl(z(\rho,\theta)d\rho^2
+(\rho^2+a^2\cos^2\theta )d\theta^2+ (\rho^2+a^2)\sin^2\theta d\phi^2
\biggr) , \label{ellips4D}
 \end{eqnarray}
  where  $v(\rho,\theta)$,
$A(\rho,\theta)$ and $z(\rho,\theta)$ are the unknown metric functions. The
metric function $A(\rho,\theta)$ is recovered from the quantities used in the
projection formalism through the equations  \be {{\partial A}\over{\partial
\rho}}=-\ {\sqrt{u}\over{v^2}} {{\partial\omega}\over{\partial\theta}}\sin\theta
, \ {{\partial A}\over{\partial\theta}}=-\ {{\rho^2+a^2}\over{v^2\sqrt{u}}}
{{\partial\omega}\over{\partial\rho}}\sin\theta , \label{Ageneral}  \ee where we
have introduced the function \be
u=u(\rho,\theta):=z(\rho,\theta){{\rho^2+a^2}\over{\rho^2+a^2\cos^2\theta}}.
\label{u}\ee In  the coordinate system of (\ref{ellips4D}), from the field
equation (\ref{fe1}) one has  \be 16\pi v^{-1}P= {R_{33}\over h_{33}}
\label{Poverv} \ee and therefore one can write
\begin{equation}H_{AB}:=R_{AB}-{R_{33}\over h_{33}}h_{AB}={1\over 2}v^{-2}\bigl[
(\partial_Av)(\partial_Bv)+(\partial_A\omega )(\partial_B\omega )\bigr]\ ,
\label{HABdef} \end{equation} where $A,B=1,2$. Notice that the left-hand side of
the above equation, i.e., the tensor $H_{AB}$ introduced here, depends merely on
the 3-dimen\-sional Riemannian metric $h_{ab}$ \cite{RaczJMP, RaczERE}. Since in
the present coordinate system the only unspecified component of $h_{ab}$ is
given by the function $z(\rho,\theta)$, it follows that the only unknown
function entering into $H_{AB}$ is $z(\rho,\theta)$. This property will be
useful in the study of the integrability conditions of the field equations,
since they will reduce to a system of equations for the single unknown function
$z(\rho,\theta)$ \cite{RaczJMP}.

Analogously to the vacuum case considered in \cite{RaczSuveges}, in studying the
above field equations it is convenient to consider separately two cases: one in
which the twist, $D_a\omega$, and the gradient of the norm of the Killing field,
$D_av$, are linearly dependent, and the other in which they are linearly
independent.

\subsection{$D_av$ and $D_a\omega$ linearly dependent}
\subsubsection{The general form of the metric}

If $D_av \not= 0$ and $D_a\omega \not= 0$ are linearly dependent, then
the functions $v$ and $\omega$ are functionally
related, so that one can write $\omega=\omega(v)$. Then using equation
(\ref{HABdef}) one obtains
\begin{equation}
H_{AB}={1\over 2}\partial_AV\partial_BV   , \label{HAB}
\end{equation}
 where $A, B= 1,2$ and we have introduced the function $V$ by
\begin{equation}
\partial_AV:=v^{-1}\sqrt{1+(d\omega/d v)^2}\partial_Av .
\end{equation}
From equation (\ref{HAB}) it follows that in any local
coordinate system
\begin{equation}
H_{11}H_{22}-H_{12}^2=0 . \label{HABcond}
\end{equation}
Furthermore, the integrability condition for the existence of a function $V$
satisfying (\ref{HAB}) can be written as
\begin{equation}
H_{11}{{\partial H_{22}}\over{\partial x^1}}-H_{12}{{\partial
H_{11}}\over{\partial x^2}}=0   .\label{intcond} \end{equation}
Equations (\ref{HABcond}) and (\ref{intcond}) are direct generalizations of the
conditions obtained for the vacuum case in \cite{RaczSuveges}.

Recall that $H_{AB}$ depends merely on the 3-dimensional metric $h_{ab}$, and
the only unknown quantity entering $H_{ab}$ is the function $z(\rho,\theta)$.
Thus, in the coordinate system ($x^1=\rho$,  $x^2=\theta$), the
equations (\ref{HABcond}) and (\ref{intcond}) represent a system of two partial
differential equations for the single unknown function $z(\rho,\theta)$. It is
straightforward to check that the choice $z=z(\rho)$ and $a=0$ is a sufficient
condition for both equations (\ref{HABcond}) and (\ref{intcond}) to be
satisfied.  It can be shown, however, that if the pressure of the fluid
satisfies $P=P(\rho)$, then this condition is not only sufficient, but also
necessary for the existence of perfect-fluid spacetimes with confocal inside
ellipsoidal symmetry having $v$ and $\omega$ functionally related.

Namely, suppose that the surfaces of constant pressure are given by the
coordinate ellipsoids $\rho=const$, that is, $P=P(\rho)$. Because of the
equation of motion (\ref{EL}), (if $\mu+P\ne 0$) this implies $v=v(\rho)$,
which, in turn, implies $\omega=\omega(\rho)$, since $v$ and $\omega$ were
required to be functionally related. In addition, by calculating the Ricci
tensor $R_{ab}$ for the 3-dimensional metric $h_{ab}$ one obtains for the single
non-diagonal component of $R_{ab}$
\be
R_{12}=- {{\rho \over{2u(\rho^2+a^2)\sin\theta}}
{{\partial u}\over{\partial\theta}}},\ee
where the function $u=u(\rho,\theta)$ is defined by equation (\ref{u}). With
this result, setting $a=1, b=2$ in the field equation (\ref{fe1}) yields
$\partial u/ \partial\theta=0$ , i.e., $u=u(\rho)$. Now using the metric
corresponding to the examined case of spacetimes, one obtains that
(\ref{HABcond}) reduces to
\be
a^3\biggl[\rho{{\partial u}\over{\partial\rho}}-2u(1-u)\biggr]\cos^2\theta+
a\biggl[\rho{{\partial u}\over{\partial\rho}}+2u(1-u)\biggr]\rho^2=0 .
\label{HABcondexact}\ee
Since $\rho$ and $\theta$ are independent coordinates, the term in the above
equation with $\cos\theta$, as well as the one without $\cos\theta$
should vanish. Excluding the solution $u=1$, which recovers Minkowski
spacetime, and the trivial solution $u=0$, one obtains that for the
perfect-fluid spacetimes under consideration the necessary and
sufficient condition for equation (\ref{HABcondexact}) to be satisfied is $a=0$.
With this condition equation (\ref{intcond}) is also identically satisfied.

On the other hand, $a=0$ and $u=u(\rho)$ imply $u=z=z(\rho)$, so the
system of equations (\ref{Ageneral}) for recovering the metric function $A$ from
the quantities used in the projection formalism now reduces to the single
equation \begin{equation}
{1\over{\sin\theta}}{d A\over{d\theta}}=-v^{-2}\rho^2z^{-1/2}
{d\omega\over{d\rho}}=:2R ,\label{Aeq} \end{equation}
where the constant $R$ could be introduced because the left-hand side
of (\ref{Aeq}) depends merely on the variable $\theta$, while the right-hand
side depends merely on $\rho$, so both sides should be equal to a constant. Then
it follows that $A=2R\cos\theta$. Therefore, using (\ref{ellips4D}), the line
element of  all stationary, axially symmetric, rigidly rotating perfect-fluid
configurations with confocal inside ellipsoidal symmetry for which the gradient
of the norm of the timelike Killing vector field and its twist are linearly
dependent can be given as
\begin{equation}  ds^2=v(\rho)(dt+2R\cos\theta
d\phi)^2-{1\over v(\rho)} \biggl(z(\rho)d\rho^2+\rho^2(d\theta^2+\sin^2\theta
d\phi^2)\biggr), \label{insideNUTlike} \end{equation}
provided that the surfaces of constant pressure are given by the $\rho=const$
ellipsoids and provided that $\mu+P\not =0$ \cite{ZsigraiJGRG11}. Equivalently,
by performing the coordinate transformation
$-v(\rho)^{-1}\rho^2\rightarrow\rho^2$ and introducing the function
$f(\rho):=-z(\rho)/v(\rho)$ one may write the metric also in the form
\begin{equation} ds^2=v(\rho)(dt+2R\cos\theta
d\phi)^2+f(\rho)d\rho^2+\rho^2(d\theta^2+\sin^2\theta d\phi^2) , \label{NUTlike}
\end{equation} which corresponds to a special case of spacetimes with co-moving
ellipsoidal symmetry in the sense of Definition 3.

The algebraic type  of the metrics (\ref{insideNUTlike}) and
(\ref{NUTlike}) is Petrov type D. Unfortunately, they are not suitable for
describing the interior of compact rotating bodies with a regular axis of
rotation. Namely, the existence of a regular rotation axis would require
$g_{\phi\phi}\rightarrow 0$ as $\theta\rightarrow 0$, for arbitrary
$\rho$. This, however, is not possible with the metrics (\ref{insideNUTlike}) or
(\ref{NUTlike}), therefore they cannot have a regular rotation axis. It turns
out, instead, that they are locally rotationally symmetric and they belong to
Class I in the Stewart-Ellis classification \cite{StewartEllis} of locally
rotationally symmetric spacetimes.

Next we show that these perfect-fluid spacetimes can be matched to an
external Taub-NUT solution, provided that there exists a $P=0$ surface. The
line-element of the Taub-NUT metric is
 \begin{equation}
ds^2=-F(r)(d\tau+2l\cos\theta
d\phi)^2+F^{-1}(r)dr^2+(r^2+l^2)(d\theta^2+\sin^2\theta d\phi^2) \ ,
\label{dsNUT} 
\end{equation}
\begin{equation}
 F(r)={{r^2-2mr-l^2}\over{r^2+l^2}}\ ,
\end{equation}
where $l$ and $m$ are constants \cite{Exact}. For the purpose of the matching we
 assume that
there exists a $\rho_s$  for which $P(\rho_s)=0$ and that on the
$\rho=\rho_s$ surface the relationships $\tau=\eta t$ ($\eta=const$) and
$r=r(\rho)$ hold. We join the two spacetimes by requiring the continuity of
the metric components at the junction surface. In addition, the components
of the extrinsic curvature tensor of the matching surface are also expected
to be equal on the two sides of the surface. Then one can express the
parameters $m$ and $l$ of the Taub-NUT metric, as well as the junction parameter
$\eta$, in terms of the functions of the interior spacetime evaluated on the
surface $P=0$. That is, there exists a matching of the examined perfect-fluid
spacetimes to the Taub-NUT metric with
 \begin{equation}
\eta^2={{\rho_s^2v_sf_s}\over{R^2v_sf_s-\rho_s^2}} \end{equation}
\begin{equation} l=\eta R \end{equation} \begin{equation}
m={1\over{2\sqrt{\rho_s^2-\eta^2R^2}}}\Bigl(\rho_s^2-2\eta^2R^2+
{{v_s\rho_s^2}\over{\alpha^2}}\Bigr)
 \end{equation}
where $v_s=v(\rho_s)$ and $f_s=f(\rho_s)$. Note that
matchings of locally rotationally symmetric spacetimes to the Taub-NUT metric
were considered in \cite{Bradleyetal} as well.

\subsubsection{Examples of exact solutions}

An interesting property of the metric (\ref{insideNUTlike}) is that if the
function $v(\rho)$ is known, then $z(\rho)$ can be obtained from
equation (\ref{HABdef}) by performing two quadratures. Then the pressure and
density of the fluid are obtained from equations (\ref{EL}) and (\ref{Poverv}),
respectively, without performing any integrations. Therefore, generating new
exact solutions of this type is, in fact, a simple exercise. Unfortunately, this
way there is no guarantee that the resulting spacetime will possess a meaningful
equation of state. Thus, instead of giving a bunch of unphysical new exact
solutions, in the following we rather list some interesting examples of
previously known perfect-fluid spacetimes which can be shown to belong to the
class given by equations (\ref{insideNUTlike}) or (\ref{NUTlike}). We wish to
emphasize, however, that the list of previously known perfect-fluid solutions
possessing confocal inside ellipsoidal symmetry presented here is by no means
meant to be complete, there might as well exist other solutions in the
literature, also belonging to the examined class.

 In \cite{Gergelyetal,Perjesetal} a spacetime originally given by Ferwagner
\cite{Ferwagner} and later rediscovered by Marklund \cite{Marklund} is
extensively studied. This spacetime represents an incompressible, rigidly
rotating, Petrov type D perfect-fluid configuration. By performing the
coordinate transformation $R\sin\chi\rightarrow \rho$, $dt\rightarrow
dt+(R^2/{\rho^2)(1-\rho^2/ R^2})^{- {1\over 2}}d\rho$
on the metric given in \cite{Gergelyetal}, one arrives at
\begin{equation}
ds^2=-{\rho^4\over R^4}(dt+2R\cos\theta d\phi)^2+
{1\over{1-\rho^2/R^2}}d\rho^2+\rho^2(d\theta^2+\sin^2\theta d\phi^2) ,
\label{dsFerwagner} \end{equation} where $R$ is a constant. The above line
element has exactly the same form as (\ref{NUTlike}), with $v(\rho)=\rho^4/R^4$
and $f(\rho)=(1-\rho^2/R^2)^{-1}$. The density and pressure of the fluid are
given by  \begin{equation} 8\pi \mu={6\over R^2}, \  8\pi P={4\over
\rho^2}-{6\over R^2} . \end{equation}

Next, it has been shown in \cite{Herlt} that if $\omega=\omega(v)$ and
$\mu+3P\ne 0$, there exists such a coordinate system $(t,x^1,x^2,\phi)$ in which
the line element of all stationary, axisymmetric, rigidly rotating perfect-fluid
spacetimes can be written as
\begin{equation}
ds^2=v(x^1)(dt^2+A(x^2)d\phi^2)\\
 +{1\over v(x^1)}\bigl\{h_{11}(x^1)(dx^1)^2
+h_{22}(x^1)[(dx^2)^2+ e ^ 2 ( x ^ 2 ) d \phi^2]\bigr\} ,
\label{dsHerlt} \end{equation}
where all possible forms of the functions $e(x^2)$ and $A(x^2)$ are listed in
\cite{Herlt}. By setting $x^1=\rho$, $x^2=\theta$, $h_{22}=\rho^2$ and with the
allowed choice $e^2=\sin^2\theta$, which implies $A=2R\cos\theta$,  one
recovers the general form (\ref{insideNUTlike}) of the studied ellipsoidal
spacetimes. An exact solution of the examined type was also given in \cite{Herlt},
which, because of its complexity, we do not list here but rather refer the
reader to equations (49)-(53) in the original article \cite{Herlt}.
Finally, in \cite{Lukacsetal} by using the 3-dimensional spin coefficient
method the authors have arrived to a line element which has, up to a coordinate
transformation,  the same general form as that given by equation
(\ref{insideNUTlike}), but they could obtain an exact non-vacuum solution only
for the dust case.

Note that for the special case of rigidly rotating dust one has $P \equiv 0$ and
$v \equiv const$. Assuming that the metric has the general form given by
(\ref{insideNUTlike}), it turns out that the only rigidly rotating dust solution
satisfying the conditions (\ref{HABcond}) and (\ref{intcond}) is given by
$z(\rho)=\rho^2/(\rho^2+v^2R^2)$. This solution is identical (up to a coordinate
transformation and up to renaming the constants) to a dust solution given in
\cite{Lukacsetal}.

\subsection{$D_av$ and $D_a\omega$ linearly independent -- an exact solution}

When $D_av$ and $D_a\omega$ are linearly independent, the general form of the metric is given by (\ref{ellips4D}). In this case the components of the tensor $H_{AB}$, defined by (\ref{HABdef}), may be considered as components of a non-singular Riemannian 2-metric. It has been demonstrated in \cite{RaczJMP} that the necessary and sufficient condition for the existence of
the functions $v(\rho,\theta)$ and $\omega(\rho,\theta)$ satisfying
(\ref{HABdef}) is that the Gaussian curvature of this 2-metric is exactly minus
one. In the present case this condition represents a non-linear partial differential equation for the function $z(\rho, \theta)$. It can be shown that the particular choice \be
z(\rho,\theta)={{\rho^2+a^2\cos^2\theta}\over{\rho^2+a^2}}{\rho^2\over{m^2-\rho^
2 }}, \label{zcosmo}\ee
where $a$ and $m$ are some constants, satisfies the mentioned
integrability condition. With this choice for $z$, the metric (\ref{ellips4D})
satisfies the full set of field equations only if the remaining metric functions
are given by
\begin{equation} v(\rho,\theta)=-C^2(\rho^2+a^2\cos^2\theta),\
A(\rho,\theta)={{a\sin^2\theta}\over{C^2(\rho^2+a^2\cos^2\theta)}},
\end{equation}
where $C$ is a constant. With these functions the metric of spacetime acquires
the remarkably simple form
\begin{eqnarray}
&ds^2=&-C^2(\rho^2+a^2\cos^2\theta) dt^2 +2a\sin^2\theta dtd\phi+ \nonumber \\
& &{1\over C^2}\biggl({{\rho^2d\rho^2}\over{(\rho^2+a^2)(m^2-\rho^2)}}+
d\theta^2+\sin\theta^2 d\phi^2 \biggr) \ .
\label{dsCosmological}
\end{eqnarray}

It is interesting to note that when the Kerr metric is recast into the form (\ref{ellips4D}) (see \cite{RaczCQG}) then the function $z(\rho,\theta)$ for the Kerr metric is given by
\begin{equation}
z_{Kerr}(\rho,\theta)={{\rho^2+a^2\cos^2\theta}\over{\rho^2+a^2}}{\rho^2\over{\rho^2-m^2}}
\end{equation}
which differs from (\ref{zcosmo}) merely by a negative sign factor.

Expression (\ref{dsCosmological}) represents an exact solution of Einstein's equations
corresponding to a stationary, axially symmetric, perfect-fluid spacetime with
confocal inside ellipsoidal symmetry, which does not belong to the class of
solutions given in the previous subsection. For $\theta\rightarrow 0$ one has
$g_{\phi\phi}\rightarrow 0$ which is compatible with the existence of a regular
rotation axis. From equation (\ref{Poverv}) one obtains the pressure of the
fluid, $P$. Then, using the equation of motion (\ref{EL}), the density of the
fluid, $\mu$, can also be determined. For the particular metric
(\ref{dsCosmological}) it turns out that $\mu=-P=C^2$, i.e., the solution can be
re-interpreted as a vacuum solution with a non-zero cosmological constant
$\Lambda=C^2$. We emphasize, however, that other perfect-fluid metrics (i.e.,
different from (\ref{dsCosmological})), satisfying the integrability conditions
for spacetimes with $D_av$ and $D_a\omega$ linearly independent and possessing
confocal inside ellipsoidal symmetry might exist as well. Proving the existence
(or non-existence) of such metrics is still an open issue.

\section{Conclusion}
In this paper a  framework has been presented which can be useful in the
general-relativistic study of ellipsoidal shapes. This framework relies on the notion of a Riemannian 3-space that can be filled in with a one-parameter congruence of concentric
ellipsoids, by the use of which one can define a conformally ellipsoidal
spacetime. Since in general relativity the description of any shape depends on
the relative motion of the observers performing the description, in
order to carry out a general-relativistic study of ellipsoidal shapes one
first has to specify a choice of preferred observers. Different choices give
rise to the definitions of, e.g., spacetimes with co-moving ellipsoidal symmetry
and spacetimes with inside ellipsoidal symmetry.

An attempt to apply the presented framework in the study of compact, rapidly rotating bodies has been made. In particular, the applicability of the framework was tested by a detailed study of stationary, axially symmetric, rigidly rotating perfect-fluid
spacetimes with confocal inside ellipsoidal symmetry. In the subcase when the
twist, $D_a\omega$, and the gradient of the norm of the time-like Killing
vector field, $D_av$, are linearly related the angular dependence of the
metric has been determined under the assumption that the surfaces of constant
pressure are given by the $\rho=const$ ``ellipsoids''. The obtained
general form of the metric (see equations (\ref{NUTlike}) and
(\ref{insideNUTlike})) contains two unspecified functions of the
coordinate $\rho$ and it represents a Petrov type D, locally rotationally
symmetric fluid configuration. Since  this metric cannot have a regular axis of
rotation, it should not be expected to describe the gravitational field of a
compact rotating body, but rather, as it has been shown, interior NUT
spacetimes. Some previously known exact solutions belonging to this class have
been recalled, and it has been indicated that new solutions of this type can be
easily constructed, although their physical significance is questionable. In the
complementary subcase  in which   $D_a\omega$ and $D_av$ are linearly
independent a possibly new exact perfect-fluid solution possessing a regular
axis of rotation has been obtained. Unfortunately, it has the equation of state
$\mu=-P=C^2$ and therefore corresponds to a vacuum configuration with a non-zero
cosmological constant.

The question whether in the latter subcase one might obtain perfect-fluid
solutions appropriate for describing the interior of rapidly rotating compact
objects with a regular rotation axis and a physically meaningful equation of
state is still an open issue. Furthermore, it would be instructive to check
whether the results obtained here still remain valid if one drops the assumption
that the surfaces of constant pressure are given by the $\rho=const$ surfaces
and assumes merely that the surface $P=0$ is given by a coordinate ellipsoid. An
additional step forward along the line of research presented here would be to
alter the choice of preferred observers and to consider, e.g., spacetimes with
co-moving instead of inside ellipsoidal symmetry. These issues are the subject
of ongoing work.

\section*{Acknowledgements}
I wish to thank I. R\'acz for fruitful discussions and his helpful comments
on the manuscript. This research has been made possible through the
post-doctoral fellowship scheme of the Japan Society for the Promotion of
Science under fellowship number P00787 and it has been also supported in part by
the Hungarian National Scientific Research Fund, OTKA, grant No. T030374.

\section*{Appendix}
\appendix
\setcounter{section}{0}
\section{Ellipsoids in flat space}

\setcounter{equation}{0}

This appendix contains a review of the properties of ellipsoids in flat
Euclidean space on which Definition 1 in the main part of this paper is based. Note that the definition given in \cite{RaczCQG} also relies on the properties of Euclidean ellipsoids listed in this Appendix. 

In \cite{Krasinski} an
analysis of a 3-dimensional Euclidean space filled with a congruence of
ellipsoids of revolution with a common center and common axis of symmetry was
given. In order to remain inside the class of axially symmetric configurations,
the following problem was left out from \cite{Krasinski} and left for further
considerations:

\medskip
 {\bf Problem I.} Make an analysis for nonsymmetric ellipsoids, analogous to
that given for symmetric ellipsoids in \cite{Krasinski}.
\medskip

To consider this problem, imagine a 3-dimensional Euclidean space,
$\bf R^3$, filled with a congruence of ellipsoids with a common center. Let one
of the semiaxes of the ellipsoids be taken as the coordinate $\rho$ in space, so
that $\rho=const$ on a fixed ellipsoid. Let $\phi$ be the angle measured around
this axis, and let $\theta$ be a third coordinate defined at will. Analogously
to the symmetric case considered in \cite{Krasinski}, we define
$(\rho,\theta,\phi)$ in terms of Cartesian coordinates $(x,y,z)$ as follows:
\begin{eqnarray} x&=&x_0(\rho)\sin\theta \cos\phi \nonumber \\
y&=&y_0(\rho)\sin\theta \sin\phi \\
z&=&\rho\cos\theta, \nonumber
\end{eqnarray}
where $x_0(\rho)$ and $y_0(\rho)$ are arbitrary functions whose values equal the
lengths of the other two semiaxes of the $\rho=const$ ellipsoid. In these
coordinates one obtains the line element $ds_E^2:=dx^2+dy^2+dz^2$ of
Euclidean space $\bf R^3$ as:
 \begin{eqnarray}
ds_E^2&=&\biggl[((x_0')^2\cos^2\phi+(y_0')^2\sin^2\phi)\sin^2\theta
+ \cos^2\theta\biggr]\cdot d\rho^2 \nonumber \\
&+&\biggl[(x_0^2\cos^2\phi+y_0^2\sin^2\phi)\cos^2\theta+
\rho^2\sin^2\theta\biggr]\cdot d\theta^2 \nonumber\\
&+&(x_0^2\sin^2\phi+y_0^2\cos^2\phi)\sin^2\theta \cdot d\phi^2 \label{dsEuclideanEllipsoidal}\\
&+&2(x_0x_0'\cos^2\phi+y_0y_0'\sin^2\phi-\rho)\cos\theta\sin\theta \cdot d\rho
d\theta \nonumber \\
&+&2(-x_0x_0'+y_0y_0')\cos\phi\sin\phi \sin^2\theta\cdot d\rho d\phi \nonumber\\
 &+&2(y_0^2-x_0^2)\cos\phi\sin\phi \cos\theta\sin\theta\cdot d\theta d\phi \nonumber
 \ , 
\end{eqnarray}
where the prime denotes derivation with respect to $\rho$.

The special case in which the congruence consists of
confocal (oblate) ellipsoids is recovered by setting  $x_0^2=\rho^2+a^2 $,
$y_0^2=\rho^2+b^2$, where $a$ and $b$ are constants.
On the other hand, for the special case of rotationally invariant ellipsoids,
we have $x_0=y_0=:g$,
which reproduces equation (3.2) of \cite{Krasinski}:
 \begin{eqnarray}
ds_E^2&=&((g')^2\sin^2\theta+\cos^2\theta)d\rho^2+
2(gg'-\rho)\sin\theta\cos\theta d\rho d\theta \nonumber\\
 &+&(g^2\cos^2\theta+\rho^2\sin^2\theta)d\theta^2+
g^2\sin^2\theta d\phi^2.
\end{eqnarray}
Finally, if one considers a congruence of rotationally invariant confocal oblate
ellipsoids, one has $g^2=\rho^2+a^2$
and the metric becomes
\begin{eqnarray}  
ds_E^2={{\rho^2+a^2\cos^2\theta}\over{\rho^2+a^2}}d\rho^2
+(\rho^2+a^2\cos^2\theta)d\theta^2+(\rho^2+a^2)\sin^2\theta d\phi^2 .
\label{dsEconfocalrotinv}
 \end{eqnarray}

Note that by letting $\rho=const$ in any of the above metrics one obtains the
metric induced on a single ellipsoidal surface with the ``radius'' $\rho$. By
using the coordinate system of (\ref{dsEuclideanEllipsoidal}) it is
straightforward to show that the extrinsic curvature tensor $\kappa_{ab}$  of
any one of the $\rho=const$ surfaces is conformal to the metric of the unit
sphere, $\gamma_{ab}$, which may be given as
$ds^2(\gamma_{ab})=d\theta^2+\sin\theta d\phi^2$. That is
 \be
\kappa_{ab}=k\gamma_{ab}, \label{EuclideanExtrinsic}
\ee
where $k$ is a conformal factor, which is a function of the coordinates chosen
to represent 3-dimensional Euclidean space. On the other hand, the intrinsic geometry of a
{\it singe surface} does not depend on the properties of the congruence in which
it is embedded in. Namely, in Euclidean space the surface of any $\rho=const$
ellipsoid is conformal to the unit sphere, which implies that  one can always
introduce such coordinates on the surface in which the metric of the surface can
be put into the form $K\gamma_{ab}$, where $K$ is again some conformal factor.
This is an intrinsic property of each $\rho=const$ ellipsoid. In this respect,
equation (\ref{EuclideanExtrinsic}) provides a connection between the
``intrinsic'' and ``extrinsic'' properties of the ellipsoidal surfaces in
Euclidean space.

\end{document}